\documentclass[onecolumn]{revtex4}

\usepackage{graphicx}
\usepackage{dcolumn}
\usepackage{bm}

\input epsf

\begin{document}

\title{\Large Generalized Bekenstein-Hawking System : Logarithmic Correction }

\author{\bf Subenoy Chakraborty\footnote{schakraborty@math.jdvu.ac.in }}

\affiliation{Department Of Mathematics, Jadavpur University, Kolkata, India.}{}



\begin{abstract}
The present work is a generalization of the recent work [arXiv.no.1206.1420] 
on the modified Hawking temperature
on the event horizon. Here the Hawking temperature is 
generalized by multiplying the modified Hawking temperature
by a variable parameter $\alpha$ representing the ratio of 
the growth rate of the apparent horizon to that of event
horizon. It is found that both the first and the generalized 
second law of thermodynamics are valid on the event horizon
for any fluid distribution. Subsequently, Bekenstein entropy 
is modified on the event horizon and thermodynamical laws are 
examined. Finally, interpretation of the parameters involved has been presented.\\


\end{abstract}

\maketitle

\section{Introduction}

In black hole physics a semi classical description shows that a
black hole behaves as a black body emitting thermal radiation with
temperature (known as Hawking temperature) and entropy(known as
Bekenstein entropy) proportional to the surface gravity at the
horizon and area of the horizon $[1,2]$ respectively. Further,this
Hawking temperature, and Bekenstein entropy are related to the
mass of the black hole through the first law of thermodynamics
$[3]$. Due to this inter relationship between the physical
parameters(namely, entropy,temperature) and the geometry of the
horizon,there is natural speculation about the inter relationship
between the black hole thermodynamics and the Einstein
field equations.A first step in this direction was put forward by
Jacobson $[4]$ who derived the Einstein field equations from the
first law of thermodynamics : $\delta Q=T dS$ for all locally
Rindler causal horizons with   $\delta Q$ and T as the energy flux
and Unruh temperature measured by an accelerated observer just
inside the horizon. Subsequently, Padmanabhan $[5]$ from the other
side was able to derive the first law of thermodynamics on the
horizon starting from Einstein equations for a general static
spherically symmetric space time.

This idea of equivalence between Einstein field equations and the
thermodynamical laws has been extended in the context of cosmology.
Usually ,universe bounded by the apparent horizon is assumed to be
a thermodynamical system with Hawking temperature and the entropy
as,
\begin{eqnarray}
T_{A}=\frac{1}{2\pi R_{A}}
\nonumber
\\
S_{A}=\frac{\pi R_{A}^{2}}{G}
\end{eqnarray}
where $R_A$ is the radius of the apparent horizon.
It was shown that the first law of thermodynamics on the apparent horizon and
the Friedmann equations are equivalent $[6]$ .Subsequntly,this equivalent idea was
extended to higher dimensional space-time namely gravity theory with Gauss-Bonnet
term and for the lovelock gravity theory $[6-8]$ . It is presumed that such a inherent
relationship between the thermodynamics at the apparent horizon and the Einstein field
equations may lead to some clue on the properties of dark energy.

Although, the cosmological event horizon does not exist in the usual standard big bang
cosmology,but in the perspective of the recent observations $[9-11]$, the universe is in an
accelerating phase dominated by dark energy($\omega_d<-1/3$) and the event horizon distinct
from the apparent horizon.By defining the entropy and temperature on the event horizon
similar to those for the apparent horizon (given above) Wang et al $[12]$ showed that both
the first and the second law of thermodynamics breakdown on the cosmological event horizon.
They justified it arguing that the first law is applicable to nearby states of local
thermodynamic equilibrium while the event horizon reflects the global features of space-time.
As a result,the thermodynamical parameters on the non-equilibrium configuration of the event horizon
may not be as simple as on the apparent horizon. Further,they speculated that the region bounded by the
apparent horizon may be taken as the Bekenstein system i.e.Bekenstein's entropy or mass bound:
$ S<2\Pi R_{E}$ and entropy or area  bound:$ S<A/4$ are satisfied 
in this region. Now due to universality of the
Bekenstein bounds and as all gravitationally stable special regions with weak self-gravity should satisfy
the above Bekenstein bounds so the corresponding thermodynamical system is termed as a Bekenstein system.
Further,due to larger radius of the event horizon than the apparent horizon ,Wang et al $[12]$ termed the
universe bounded by the event horizon as a non-Bekenstein system.

In recent past there were a series of works $[13-16]$ 
investigating the validity of the generalized second law of
thermodynamics of the universe bounded by the event horizon 
for Einstein gravity $[13,14]$ and in other gravity theories
$[13-15]$ and for different fluid systems $[13,14,16]$(including dark energy $[14,16]$). 
In these works the validity of the
first law of thermodynamics on the event horizon was assumed and it was 
possible to show the validity of the generalized
second law of thermodynamics with some reasonable restrictions. 
However, validity of the first law of thermodynamics on
the event horizon was still in a question mark. Very recently, 
the author $[17]$ is able to show that the first law of
thermodynamics is satisfied on the event horizon with a modified 
Hawking temperature for two specific examples of single
DE fluids. The present work is a further, extension of it. 
Here by generalizing the Hawking temperature, or modifying 
Bekenstein entropy
it is possible to show that both the first and the 
generalized second law of thermodynamics(GSLT) are always 
satisfied on the event horizon. The paper
is organized as follows : Section 2 deals with basic equations 
related to earlier works . Thermodynamical laws with generalized Hawking temperature
and modified Bekenstein entropy has been studied respectively 
in section 3 and in section 4 .Interpretation of the parameters involved in generalized
Hawking temperature and modified Bekenstein entropy has been 
analyzed in section 5 . Finally , summary of the work and possible 
conclusions are presented in section 6.

\section{Basic Equations and Earlier Works}
The homogeneous and isotropic FRW model of the Universe can locally be expressed by the metric as
\begin{equation}
ds^{2} = h_{ij}(x^{i}) dx^{i}dx^{j} +R^{2} d\Omega_{2}^{2}
\end{equation}
where i,j can take values 0 and 1,the two dimensional metric tensor $h_{ij}$, known as normal metric is given by
\begin{equation}
h_{ij} = diag(-1,a^{2}/1-kr^{2})
\end{equation}
with $x^{i}$ being associated co-ordinates $( x^{0}= t,x^{1}= r)$. $R=ar$ is the area radius and is considered as a scalar field
in the normal 2D space. Another relevant scalar quantity on this normal space is
\begin{equation}
\chi(x)= h^{ij}(x)\partial_{i}R\partial_{j}R = 1-(H^{2}+k/a^{2})R^{2}
\end{equation}
where $k=0,\pm 1$ stands for flat,closed or open model of the Universe.The Friedmann equations are
\begin{equation}
H^{2} + k/a^{2} =\frac{8\Pi G\rho}{3}
\end{equation}
and
\begin{equation}
\dot{H}- k/a^{2} = -4\Pi G(\rho + p)
\end{equation}
where the energy density $\rho$ and the thermodynamic pressure p of the matter distribution obey the conservation relation
\begin{equation}
\dot{\rho} + 3H(\rho + p) = 0
\end{equation}
Usually, the apparent horizon is defined at the vanishing of the scalar i.e. $\chi(x) = 0$,which gives
\begin{equation}
R_{A} = \frac{1}{\sqrt{H^{2}+ k/a^{2}}}
\end{equation}
Now the surface gravity on the apparent horizon is defined as
\begin{equation}
\kappa_{A = -\frac{1}{2}}\frac{\partial\chi}{\partial R}\|_{R=R_{A}} =\frac{1}{R_{A}}
\end{equation}
So the usual Hawking temperature on the apparent horizon is given by(as in equation(1))
\begin{equation}
T_{A} = \frac{\|\kappa_{A}\|}{2\Pi} =\frac{1}{2\Pi R_{A}}
\end{equation}
It has been shown by Wang etal $[12]$ and others $[6,7,18]$ that Universe bounded by the apparent horizon (with parameters given
by equation(1)) is a thermodynamical system satisfying both the first and the second law of thermodynamics  not only in
Einstein gravity but also in any other gravity theory and also for baryonic as well as for exotic matter.

On the other hand,the difficulty starts from the very definition of the event horizon.The infinite integral in the definition
\begin{equation}
R_{E} =a\int_{t}^{\infty} \frac{dt}{a}
\end{equation}
converges only if $ a\sim t^{\alpha}$ with $\alpha>1$ i.e. the event horizon does not exist in the decelerating phase,it has
only relevance in the present accelerating era. In the literature ,the Hawking temperature on the event horizon is usually
taken similar to the apparent horizon (replacing $R_{A }$ by $R_{E }$) as (see eq.(10))
\begin{equation}
T_{E} =\frac{1}{2\Pi R_{E}}
\end{equation}
This choice is also supported from the measurement of the temperature by a freely falling detector in a de-Sitter background(where both the
horizons coincide) using quantum field theory $[19]$ . But unfortunately ,with this choice of temperature and the entropy in the form of
Bekenstein i.e.

\begin{equation}
T_{E} = \frac{1}{2\Pi R_{E}},~~~~~ S_{E} =\frac{\Pi R_{E}^{2}}{G}
\end{equation}

the universe bounded by the event horizon is not a realistic thermodynamical system as both the thermodynamical laws fail to hold there $[12]$.\\

Recently,the surface gravity on the event horizon is defined similar to that on the apparent horizon (see eq.(9))as $[17]$

\begin{equation}
\kappa_{E} =-\frac{1}{2} \frac{\partial\chi}{\partial R}|_{R = R_{E}} =\frac{R_{E}}{R_{A}^{2}}
\end{equation}

and as a result the modified Hawking temperature on the event horizon becomes

\begin{equation}
T_{E}^{m} = \frac{\|\kappa_{E}\|}{2\Pi} = \frac{R_{E}}{2\Pi R_{A}^{2}}
\end{equation}

which for flat $FRW$ model (i.e. $k=0$) becomes
\begin{equation}
T_{E}= \frac{H^{2}R_{E}}{2\Pi}
\end{equation}
As the two horizons are related by the inequality

\begin{equation}
R_{A} < R_{E}
\end{equation}

so we always have

\begin{equation}
T_{A} < T_{E}
\end{equation}

Using this modified Hawking temperature the author $[17]$ is able to show the validity of the first law of thermodynamics
on the event horizon for two specific single fluid DE model.

\section{Generalized Hawking Temperature and Thermodynamical Laws}

In this section to proceed further for a general prescription, 
we start with a generalization of the modified Hawking temperature
in the form

\begin{equation}
T_{E}^{g}= \frac{\alpha R_{E}}{2\Pi R_{A}^{2}}
\end{equation}

where the dimensionless parameter $\alpha$ is to be determined so that $\alpha = 1$ on the apparent horizon.

The amount of energy flux across a horizon  within the time interval $dt$ is $[6,20]$

\begin{equation}
-dE_{h} = 4\Pi R_{h}^{2}T_{ab}k^{a}k^{b}dt,
\end{equation}

with $k^{a}$, a null vector. So for the event horizon we get

\begin{equation}
-dE = 4\Pi R_{E}^{3}H(\rho+p)dt.
\end{equation}

Now using the Einstein field equation (6) and the definition 
of the apparent horizon (i.e. eq(8)),the above expression for
energy flux simplifies to

\begin{equation}
-dE = (\frac{R_{E}}{R_{A}})^{3} \frac{\dot{R_{A}}}{G}dt.
\end{equation}

From the Bekenstein's entropy -area relation (see eq(13)) we have

\begin{equation}
T_{E} dS_{E} = \alpha(\frac{R_{E}}{R_{A}})^{2} \frac{\dot{R_{E}}}{G}dt.
\end{equation}

Hence for the validity of the first law of thermodynamics i.e.

\begin{equation}
-dE = dQ = T_{E} dS_{E},
\end{equation}

we have

\begin{equation}
\alpha = \frac{\dot{R_{A}}/R_{A}}{\dot{R_{E}}/R_{E}}.
\end{equation}

Thus reciprocal of $\alpha$ gives the relative growth rate of the radius of the event horizon to the apparent horizon.

For the generalized second law of thermodynamics ,we start with the Gibb's law $[12,21]$ to find the entropy variation of the bounded fluid
distribution:

\begin{equation}
T_{f} dS_{f} = dE + p dV,
\end{equation}

where $T_{f}$ and $S_{f}$ are the temperature and entropy of the given fluid distribution respectively,$V= 4\Pi R_{E}^{3}/3$
and $E= \rho V$ .The above equation explicitly takes the form

\begin{equation}
T_{f} dS_{f} = 4\Pi R_{E}^{2}(\rho + p)(\dot{R_{E}}- H R_{E})dt
\end{equation}

Also using the first law (i.e. eq(24)) we have from (21)

\begin{equation}
T_{E} dS_{E} = 4\Pi R_{E}^{3} H(\rho +p)dt
\end{equation}

Now for equilibrium distribution,we assume $T_{f} = T_{E}^{g}$ i.e. the inside matter has the same temperature as the
bounding surface and we obtain

\begin{equation}
T_{E} dS_{T} = 4\Pi R_{E}^{2}(\rho + p)\dot{R_{E}} dt
\end{equation}

with $S_{T} = S_{E} + S_{f} $,the total entropy of the universal system.
Again using the Einstein field equation(6),conservation relation (7) and the equation (8) we have on simplification

\begin{equation}
T_{E}^{g}\frac{dS_{T}}{dt} =( \frac{R_{E}}{R_{A}})^{2} \frac{\dot{R_{A}}\dot{R_{E}}}{GHR_{A}}
\end{equation}

Now using the generalized Hawking temperature (19) the time variation of the total entropy becomes

\begin{equation}
\frac{dS_{T}}{dt} = \frac{2\Pi}{GH} \dot{R_{E}}^{2}
\end{equation}

which is positive definite for expanding Universe and hence the generalized second law of thermodynamics always
holds on the event horizon.

\section{Modified Bekenstein Entropy and Thermodynamical Laws}
In the previous section we have generalized the Hawking temperature, keeping the Bekenstein entropy - area relation unchanged and we are able to
show the validity of both the first law of thermodynamics and GSLT on the event horizon, irrespective of any fluid distribution and we may termed
universe bounded by the event horizon as a generalized Bekenstein system. However, it is possible to have two other modifications of entropy and
temperature on the event horizon as\\
a) $S_{E}^{(m)} = \beta S_{E}^{(B)} , T_{E} = T_{E}^{(m)}$  and b) $S_{E}^{(m)} = \delta S_{E}^{(B)} , T_{E} = \frac{1}{\delta} T_{E}^{(m)}$.\\
 We shall now examine the validity of the thermodynamical laws for these choices :\\
\textbf{a)}  $S_{E}^{(m)} = \beta S_{E}^{(B)}$  and  $T_{E} = T_{E}^{(m)}$\\
Here $S_{E}^{(m)}$ and $S_{E}^{(B)}$ are respectively the modified entropy and the usual Bekenstein entropy on the event horizon, $T_{E}^{(m)}$
is the modified Hawking temperature on the event horizon (given in equation (15) or (16)) and $\beta$ is a parameter having value unity on the
apparent horizon. Then as before from the validity of the Clausius relation $\beta$ can be determined as

\begin{equation}
\beta = \frac{2}{R_{E}^{2}}\int R_{E}^{2}\frac{d R_{A}}{R_{A}}
\end{equation}

Thus for this choice of $\beta$ the above modified entropy and modified Hawking temperature satisfy first law of thermodynamics on the event
horizon. Now we shall examine the validity of the generalized second law of thermodynamics (GSLT) on the event horizon for this choice of entropy
and temperature on the horizon. Proceeding as before (assuming the temperature of the inside fluid is same as modified Hawking temperature for
thermodynamical equilibrium ) we have

\begin{equation}
\frac{dS_{T}}{dt} = \frac{2 \Pi}{G H}(\frac{R_{E}}{R_{A}}) \dot{R_{A}}\dot{R_{E}}
\end{equation}

Thus validity of GSLT  depends on the evolution of the two horizons (apparent and event)- if both the horizons increase or decrease simultaneously the GSLT is always satisfied. However, as long as weak energy condition (WEC) is satisfied $\dot{R_{A}} > 0$ and $\dot{R_{E}} > 0$ if $R_{E} > R_{A}$ and GSLT is satisfied. But if WEC is violated then $\dot{R_{A}}< 0$ and GSLT will be satisfied only if $\dot{R_{E}}< 0$ i.e. $R_{E} < R_{A}$, which may be
possible only in phantom era. Hence for this choice of entropy and temperature on the event horizon GSLT is always satisfied as long as WEC is satisfied and when WEC is violated then GSLT will be valid if $R_{E} < R_{A}$. Further,it should be noted that if we choose the temperature on the event horizon as the generalized Hawking temperature (i.e. $T_{E}^{g}$) then $\beta$ turns out to be unity i.e. we get back to the previous generalized Bekenstein system (in section 3).\\
\textbf{b)}  $S_{E}^{(m)} = \delta S_{E}^{(B)} , T_{E} = \frac{1}{\delta} T_{E}^{(m)}$\\
As before the parameter $\delta$ should be unity on the apparent horizon to match with the Bekenstein system. Again for the validity of the first law of thermodynamics (i.e. Clausius relation) $\delta$ turns out to be $R_{A}^{2}/R_{E}^{2}$ and as a result the entropy on the event horizon becomes constant (to that at the apparent horizon). So this choice of entropy - temperature is not of much physical interest.

\section{Interpretation of the parameters $\alpha$ and $\beta$}
\textbf{I)}  $\alpha$- parameter\\

In this section We shall try to make some implications of the factor $\alpha$ . Note that $\alpha$ can be termed as the ratio of the
expansion rate of the two horizons. If we compare the expansion rate of the expanding matter with that for both
the horizons, we have

\begin{equation}
\frac{\dot{R_{E}}}{R_{E}} - H = -\frac{1}{R_{E}}        , \frac{\dot{R_{A}}}{R_{A}} - H = (\frac{3\omega + 1}{2})H
\end{equation}

where the matter in the universe is chosen as a barotropic fluid with equation of state :$p = \omega \rho$ . As event horizon
exists only for accelerating phase so $\omega< -\frac{1}{3}$ . Hence both the horizons expand slower than comoving. So expansion rate of both the
horizons coincide (i.e. $\alpha = 1$)when

 \begin{equation}
 H R_{E} = -\frac{2}{(3\omega + 1)}
 \end{equation}

 Before proceeding further , we present a comparative characterization of the two horizons in table I.

\begin{table}
\begin{center}

\caption{\bf  A Comparative Study of The Horizons}
\centering

\begin{tabular}{|c|c|c|c|c|}
\hline
{\bf Horizon}& {\bf Location or Definition} & {\bf Causal Character} & {\bf Velocity} & {\bf Acceleration}\\[0.3ex]
\hline

Apparent

&
$R_{A} = $

&
Time like if $-1<\omega<\frac{1}{3}$,
&
$4\Pi HR_{A}^{2} (\rho + p) =$
&
$\frac{9H}{2}(1 + \frac{p}{\rho})$\\

Horizon
&
$\frac{1}{\sqrt{H^{2}+\frac{\kappa}{a^{2}}}}$
&
Null if $\omega = -1 or \frac{1}{3}$,
Space like if $\omega<-1 or \omega>\frac{1}{3}$.

&
$ \frac{3}{2}(1 + \frac{p}{\rho})$

&
$(\frac{p}{\rho}-\frac{\dot{p}}{\dot{\rho}})$\\
\hline

Event Horizon

&
$R_{E} = a\int_{t}^{\infty}\frac{dt'}{a(t')}$

&
Null

&
$HR_{E} - 1$

&
$-H(1 + qHR_{E})$\\
\hline

\end{tabular}
\end{center}
\end{table}

We shall now try to estimate the parameter $\alpha$ for some known fluid system :\\

\textbf{a) Perfect fluid with constant equation of state $\omega (< -\frac{1}{3})$}:\\

For flat FRW model , the cosmological solution is

$$a(t) = a_{0} t^{[\frac{2}{3(\omega +1)}]} $$

i.e.

\begin{equation}
 H(t) = \frac{2}{3(\omega + 1)t}
\end{equation}

and
\begin{equation}
R_{E}(t) = -\frac{3(1+\omega )}{(1+ 3\omega )} t.
\end{equation}

Hence

\begin{equation}
H R_{E} = -\frac{2}{(1+ 3\omega )}
\end{equation}

i.e. eq.$(33)$ is identically satisfied for all $\omega (< -\frac{1}{3})$. So for perfect fluid with constant equation of
state $(< -\frac{1}{3})$ we always have $\alpha = 1$ and hence the expansion rate of both the horizons are identical throughout the evolution. Thus the Universe bounded by the event horizon with modified /generalized Hawking temperature
( given by equation (15)/ (19)) is a Bekenstein system and it supports the results in ref $[17]$.\\

\textbf{b Interacting holographic dark energy fluid :}\\

We shall now study interacting  holographic dark energy (HDE) model consists of dark matter in the form of dust (of energy
density $\rho_{m}$ ) and HDE in the form of perfect fluid : $p_{d} = \omega_{d} \rho_{d}$. The interaction between them is
chosen as $3b^{2} H(\rho_{m} +\rho_{d})$ with $b^{2}$ the coupling constant. If $R_{E}$ is taken as the I.R. cut off then the radius
of the event horizon ($ R_{E}$ ) and the equation of state  parameter $\omega_{d}$ are given by $[22]$

\begin{equation}
R_{E} = \frac{c}{\sqrt{\Omega_{d}}H}
\end{equation}

and

\begin{equation}
\omega_{d} = -\frac{1}{3} - \frac{2\sqrt{\Omega_{d}}}{3c} -\frac{b^{2}}{\Omega_{d}}
\end{equation}

where $\Omega_{d} = \rho_{d}/(3H^{2})$ is the density parameter for dark energy and the dimensionless parameter 'c'
carries the uncertainties of the theory and is assumed to be constant. In this case equation (33) modifies as

\begin{equation}
H R_{E} = -\frac{2}{(1+ 3\omega_{t})}
\end{equation}

with

\begin{equation}
\omega_{t} = \frac{p_{d}}{(\rho_{m}+ \rho_{d})} = \omega_{d}\Omega_{d}.
\end{equation}

We shall now examine whether for this model relation (39) is satisfied or not. Using relations (37),(38) and (40) in equation (39)
we obtain a cubic equation in $x(=\sqrt{\Omega_{d}})$

\begin{equation}
2x^{3 }+ cx^{2} - 2x - (1- b^{2})c = 0.
\end{equation}

This cubic equation has a positive root ($x_{p}$) if $b^{2}< 1$ (the other two roots are either both negative or a pair of complex
congugate). In the following table we present the value of x for different choices of b and c within observational bound:\\

\begin{table}
\begin{center}

\caption{\bf  Value of x for different values of c and $b^{2}$ from eq. (41)}
\centering

\begin{tabular}{|c|c|c|c|}
\hline
{\bf c}& {\bf $b^{2}$} & {\bf $\Omega_{d}$} & {\bf x }\\[0.3ex]
\hline

0.7 & 0.92 & 0.73 & 0.85\\
0.8 & 0.84 & 0.73 & 0.85\\
0.82& 0.91 & 0.70 & 0.84\\
0.76 & 0.8 & 0.76 & 0.87\\
\hline

\end{tabular}
\end{center}
\end{table}
The table shows that for interacting DE fluids it is possible to have identical expansion rate (i.e.$ \alpha = 1$ ) for both the
horizons within observational limit of $\Omega_{d}$ and c.\\
\textbf{II)} $\beta$- parameter\\
To interpret the parameter $\beta$ we consider thermal fluctuation of the apparent horizon so that area changes by an infinitesimal amount
i.e. $A_{a}^{(m)} = A_{a} + \epsilon$, then entropy and temperature of the modified apparent horizon can be written as (from the choice (a))
$S_{a}^{(m)} = \beta S_{a}^{B}, T_{a} = T_{a}^{(m)}$.\\
Now the modified radius of the apparent horizon is related to the original radius as (in the first approximation)

\begin{equation}
R_{a}^{(m)} = R_{a} + \frac{\epsilon\prime}{2 \Pi R_{a}},~~~~~~~ \epsilon\prime = \frac{\epsilon}{4}
\end{equation}

and $\beta$ approximates to

 \begin{equation}
 \beta = 1- \frac{\epsilon\prime}{\Pi R_{a}^{2}} + \frac{2\epsilon\prime}{\Pi R_{a}^{2}} lnR_{a}
 \end{equation}

Hence we have

\begin{equation}
S_{a}^{(m)} = S_{a}^{(B)} + \frac{2\epsilon\prime}{G} ln R_{a}
\end{equation}

and

\begin{equation}
T_{a} = \frac{1}{2 \Pi R_{a}} + \frac{\epsilon\prime}{2 \Pi R_{a}^{3}}
\end{equation}

Thus there is a logarithmic  correction to the Bekenstein entropy  and the Hawking temperature is corrected by a term proportional to $R_{a}^{-3}$ due to this thermal fluctuation. However, if we consider the infinitesimal change in the radius of the apparent horizon due to the thermal fluctuation
i.e. $R_{a}^{(m)} = R_{a} + \epsilon$ then modified entropy and temperature on the horizon becomes

\begin{equation}
S_{a}^{(m)} = S_{a}^{(B)} + \frac{4\Pi \epsilon R_{a}}{G} ,~~~~~ T_{a} = T_{a}^{H} + \frac{\epsilon}{2 \Pi R_{a}^{2}}
\end{equation}

i.e. correction to Bekenstein entropy is proportional to the radius of the horizon and that of the temperature is proportional to the inverse square
of the radius.

\section{Summary and Concluding Remarks}

In this work we have studied thermodynamical laws on the event horizon for the following three choices of entropy and temperature on the event horizon:\\
1) $S_{E} = S_{E}^{(B)}, T_{E} = T_{E}^{(g)}$,~~~~ 2) $S_{E} = \beta S_{E}^{(B)}, T_{E} = T_{E}^{(m)}$, ~~~~3) $S_{E} =\delta S_{E}^{(B)}, T_{E} = \frac{1}{\delta}T_{E}^{(m)}$,\\
where $S_{E}^{(B)}$ and $T_{E}^{(m)}$ are respectively the usual Bekenstein entropy and modified Hawking temperature (given in eq.(15) or (16)) and the parameters $\alpha$, $\beta$ and $\delta$ are evaluated using Clausius relation. It is found that the parameters take value unity on the apparent horizon so that all the three choices reduce to Bekenstein- Hawking system on the apparent horizon. However, for the 3rd choice the entropy on the event horizon turns out to be constant (equal to that on the apparent horizon) and hence it is not of much physical interest. So we have not discussed it further. On the other hand , both the thermodynamical laws hold on the event horizon unconditionally for any fluid distribution for the first choice while for the second choice of entropy and temperature on the event horizon we must have $R_{E} < R_{A}$ in the phantom era for the validity of the GSLT. Hence we call universe bounded by the event horizon (for the above two choices of entropy and temperature) as a generalized Bekenstein- Hawking thermodynamical system. Also some interpretations of the parameters $\alpha$ and $\beta$  has been presented in section 5. Lastly, if the present model of the universal thermodynamics is in thermal equilibrium with CMB photons then the temperature of the horizon must coincide with the CMB temperature ($\simeq 2.73 K$) today[23,24]. Now restoring the dimension, the temperature of the event horizon (see eq.(19)) can be written as (in Kelvin)[24]
\begin{equation}
T_{E}^{(g)} = \frac{(1+q)H^{3}R_{E}^{3}}{HR_{E}-1}\frac{1}{2\pi R_{E}}(\frac{\hbar c}{k_{B}})
\end{equation}
where, $q=-(1 +\dot{H}/H^{2})$ is the usual deceleration parameter, $\hbar = 1.05\times 10^{-27}$erg-sec, $c=3\times 10^{10}$cm/sec and $k_{B}=1.38\times 10^{-16}$erg/K are respectively the Planck's constant, speed of light and Boltzmann's constant.

In particular, if we choose the cosmic fluid as holographic dark energy, then from Eq. (39) we get 
$$HR_{E}=c/\sqrt{\Omega _{d}}$$
and we have
\begin{equation}
T_{E}^{(g)}=\frac{(1+q)c^{3}}{\Omega _{d}(c-\sqrt{\Omega _{d}})}\frac{0.23}{2\pi R_{E}} K
\end{equation}
Now using the observed values of $c, \omega _{d}$, $q$ and choosing $R_{E}$ appropriately, it is always possible to match $T_{E}^{(g)}$ with the temperature of CMB photons.
Finally, the conclusions are presented below as point wise:\\

\textbf{I.} The Universe bounded by the event horizon (generalized Bekenstein-Hawking system) is a realistic thermodynamical system where both the thermodynamical laws hold for any matter system within it.\\

\textbf{II.} In deriving the thermodynamical laws we have used the second Friedmann equation (6) and the energy conservation relation (7) . On the other way assuming the first law of thermodynamics it is possible to derive the Einstein field equations. So we may conclude that the first law of thermodynamics and the Einstein field equations are equivalent (i.e. one can be derived  from the other)on the event horizon irrespective of any fluid distribution.\\

\textbf{III.} The generalized Bekenstein-Hawking system  i.e. universe bounded by the event horizon supports the recent observations i.e. results of the present work are compatible (qualitatively) to the present observed data.\\

\textbf{IV.} If due to some thermal fluctuation the apparent horizon is modified so that its area changes infinitesimally then upto first order of approximation the Bekenstein entropy is corrected by a logarithmic term and the correction to Hawking temperature is proportional to the inverse cube of the radius of the apparent horizon.\\

\textbf{V.} The horizon temperature can be in thermal equilibrium with CMB photons by appropriate choice of the parameters involved. 

For future work one may consider the following issues :\\

$i)$ The validity of the thermodynamical laws on the event horizon for other gravity theories.\\

$ii)$ Is this generalized Hawking temperature or the modified Bekenstein  valid for other horizons(if exists) of the Universal thermodynamical  system?\\

$iii)$ Further, physical and geometrical implication of the parameters $\alpha$ and $\beta $ may be interesting.\\

$iv)$ Is the present generalized Bekenstein- Hawking  system i.e., $S_{E} = S_{E}^{(B)}, T_{E} = T_{E}^{(g)}$ or $S_{E} = \beta S_{E}^{(B)}, T_{E} = T_{E}^{(m)}$ or some other modified version on the event horizon physically more realistic ?\\

  \textbf{Acknowledgement} :The work is done during a visit to IUCAA under its associateship programme. The author is thankful to IUCAA for warm hospitality and facilities at its Library. The author is thankful to UGC-DRS programme in the Dept. of Mathematics, Jadavpur University.\\

 \section{References}

 $[1]$ S.W.Hawking, \textit{Commun.Math.Phys.}\textbf{43},199 (1975).\\

 $[2]$ J.D.Bekenstein ,\textit{Phys. Rev.D} \textbf{7},2333 (1973).\\

 $[3]$ J.M.Bardeen,B.Carter and S.W.Hawking, \textit{Commun.Math.Phys.} \textbf{31}, 161 (1973).\\

 $[4]$ T.Jacobson,\textit{ Phys.Rev.Lett.}\textbf{75},1260 (1995).\\

 $[5]$ T.Padmanabhan,\textit{ Class.Quantum Grav.} \textbf{19},5387 (2002);\it{ Phys. Rept.}\textbf{406},49 (2005).\\

 $[6]$ R.G.Cai and S.P.Kim,\textit{J.High Energy Phys.}JHEP \textbf{02},050 (2005).\\

 $[7]$ M.Akbar and R.G.Cai, \textit{Phys.Lett.B} \textbf{635},7 (2006); A.Paranjape, S.Sarkar and T.Padmanavan, \textit{Phys.Rev.D} \textbf{74},104015 (2006).\\

 $[8]$ C.Lanczos,\textit{Ann.Math.}\textbf{39},842 (1938).\\

 $[9]$ A.G.Riess et al.\textit{Astron.J.}\textbf{116},1009 (1998); S.Perlmutter et al.\textit{Astrophys.J.}\textbf{517},565 (1999).\\

 $[10]$ D.N.Spergel et al.\textit{Astrophys.J.Suppl.Ser.}\textbf{148},175 (2003):\textbf{170},377 (2007).\\

 $[11]$ M.Tegmark et al. \textit{Phys.Rev.D}\textbf{69},103501 (2004); D.J.Eisenstein et al.\textit{Astrophys.J}\textbf{633},560 (2005).\\

 $[12]$ B.Wang ,Y.Gong and E.Abdalla,\textit{Phys.Rev.D}\textbf{74},083520 (2006).\\

 $[13]$ N.Mazumdar and S.Chakraborty,\textit{Class.Quantum Grav.}\textbf{26},195016 (2009).\\

 $[14]$ N.Mazumdar and S.Chakraborty,\textit{Gen.Relt.Grav.}\textbf{42},813 (2010).\\

 $[15]$ N.Mazumdar and S.Chakraborty,\textit{Eur.Phys.J.C.}\textbf{70},329 (2010);J.Dutta and S.Chakraborty,\textit{Gen.Relt.Grav.}\textbf{42},
        1863 (2010).\\

 $[16]$ S.Chakraborty, N.Mazumder and R.Biswas,\textit{Eur.Phys.Lett.}\textbf{91},4007 (2010); \textit{Gen.Relt.Grav.}\textbf{43},1827 (2011).\\

 $[17]$ S.Chakraborty,\textit{Phys.Lett.B}\textbf{718},276 (2012).(arXiv no.1206.1420)\\

 $[18]$ E.N.Saridakis and M.R.Setare,\textit{Phys.Lett.B}\textbf{670},01 (2008); E.N.Saridakis,\textit{Phys. Lett.B}\textbf{661},335(2008)\\

 $[19]$ G.W.Gibbons and S.W.Hawking,\textit{Phys.Rev.D}\textbf{15},2738 (1977).\\

 $[20]$ R.S.Bousso,\textit{Phys.Rev.D}\textit{71},064024 (2005).\\

 $[21]$ G.Izquierdo and D.Pavan,\textit{Phys.Lett.B}\textbf{633},420 (2006).\\

 $[22]$ B.Wang, Y.Gong and E. Abdalla,\textit{Phys.Lett.B}\textbf{624},141 (2005).\\

 $[23]$ E. Komatsu et. al., \textit{Five year Wilkinson Microwave Anisotropy Probe (WMAP) Observations: 
 Cosmological Interpretation} \textit{Astrophys. J. Suppl.} \textbf{180}, 330 (2009).\\
 
 $[24]$ C. Gao, \textit{Entropy}, \textbf{14}, 1296 (2012).

\end{document}